\documentclass[12pt,aps,showpacs,superscriptaddress,footinbib,preprint,noshowpacs]{revtex4-1}
\usepackage{color}
\usepackage{graphicx}
\usepackage{amsmath}
\usepackage{xcolor}
\usepackage{comment}
\usepackage{amssymb}
\usepackage{hyperref}

\usepackage{subfigure}

\begin{document}
\title{Inferring global dynamics from local structure in liquid electrolytes}

\begin{abstract}
Ion transport in concentrated electrolytes plays a fundamental role in electrochemical systems such as lithium ion batteries. Nonetheless, the mechanism of transport amid strong ion-ion interactions remains enigmatic. A key question is whether the dynamics of ion transport can be predicted by the local static structure alone, and if so what are the key structural motifs that determine transport. In this paper, we show that machine learning can successfully decompose global conductivity into the spatio-temporal average of local, instantaneous ionic contributions, and relate this ``local molar conductivity" field to the local ionic environment. Our machine learning model accurately predicts the molar conductivity of electrolyte systems that were not part of the training set, suggesting that the dynamics of ion transport is predictable from local static structure. Further, through analysing this machine-learned local conductivity field, we observe that fluctuations in local conductivity at high concentration are negatively correlated with total molar conductivity. Surprisingly, these fluctuations arise due to a long tail distribution of low conductivity ions, rather than distinct ion pairs, and are spatially correlated through both like- and unlike-charge interactions. More broadly, our approach shows how machine learning can aid the understanding of complex soft matter systems, by learning a function that attributes global collective properties to local, atomistic contributions.
\end{abstract}

\author{Penelope K. Jones}
\email{pj321@cam.ac.uk}
\affiliation{Department of Physics, University of Cambridge, CB3 0HE, Cambridge, United Kingdom}

\author{Kara D. Fong}
\affiliation{Department of Chemical and Biomolecular Engineering, University of California, Berkeley, CA 94720, USA; Energy Storage and Distributed Resources Division, Lawrence Berkeley National Laboratory, Berkeley, CA 94720, USA}

\author{Kristin A. Persson}
\affiliation{Energy Storage and Distributed Resources Division, Lawrence Berkeley National Laboratory, Berkeley, CA 94720, USA; Department of Materials Science and Engineering, University of California, Berkeley, CA 94720, USA}

\author{Alpha A. Lee}
\email{aal44@cam.ac.uk}
\affiliation{Department of Physics, University of Cambridge, CB3 0HE, Cambridge, United Kingdom}

\maketitle

The ion transport properties of electrolytes directly influence the performance of electrochemical systems such as  energy storage devices and biological systems \cite{doi:10.1021/acsenergylett.7b00792, rubinstein2012polyelectrolytes, https://doi.org/10.1002/celc.201901627}. In ideal dilute solutions, the ion conductivity is well described using the Nernst-Einstein relation. In non-dilute systems, inter-ion interactions become significant and cause several counter-intuitive observations including negative cation transference numbers \cite{pesko2017negative, PhysRevLett.122.136001, schonhoff_2018, doi:10.1021/acs.jpclett.9b00798} and unexpectedly low ion conductivities \cite{doi:10.1021/acs.jpcb.8b09439, doi:10.1021/acs.jpcb.0c02544}. 

\begin{figure}[ht]
\centering
\includegraphics[width=0.9\linewidth, trim={0cm, 26cm, 0cm, 8cm}, clip]{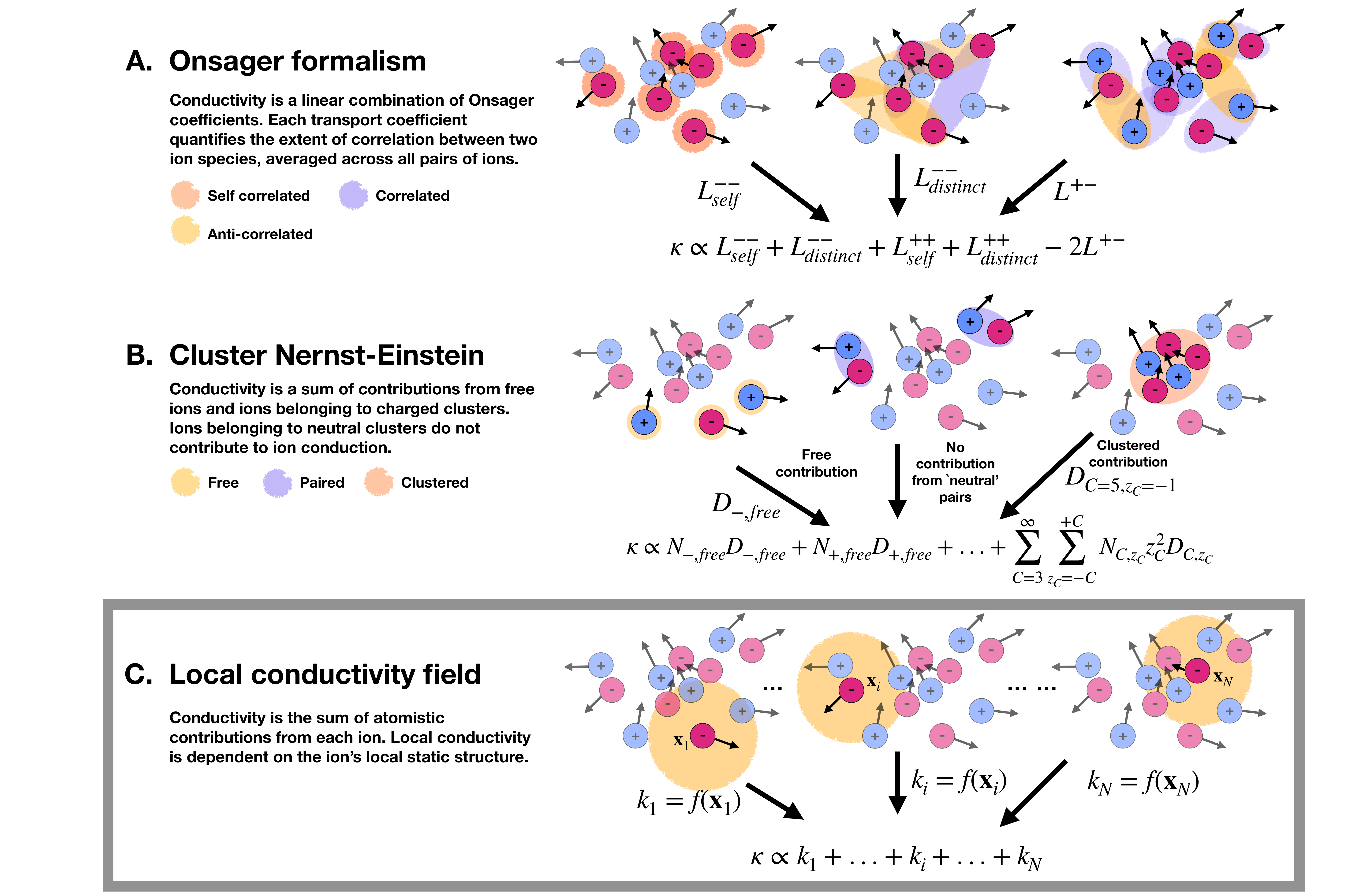}
\caption{\textbf{Schematic of our approach.} We compare our approach to the Onsager formalism \cite{fong2020onsager} and the cluster Nernst-Einstein methodology \cite{PhysRevLett.122.136001}. The Onsager schematic has been adapted from \cite{fong2020onsager}.}
\label{fig:schematic}
\end{figure}

These important and poorly understood phenomena may be rigorously captured via the Onsager transport formalism (Figure \ref{fig:schematic}A) \cite{https://doi.org/10.1002/aic.17091,doi:10.1021/acs.macromol.0c02545,PhysRev.38.2265,PhysRev.37.405}. This framework uses Green-Kubo relations and molecular simulations to compute the Onsager transport coefficients ($L^{ij}$), which map from microscopic correlations in ion motion to macroscopic transport properties such as the conductivity. Notably, these Onsager transport coefficients are global system quantities which lack spatial resolution. For example, the $L^{+-}$ transport coefficient capturing cation-anion correlations is computed based on the velocity correlation functions of every pairwise combination of cation and anion, regardless of their separation in physical space. Thus, while knowledge of $L^{+-}$ gives information on the overall importance of cation-anion correlations to transport in the system, it does not give insight as to which short- or long-ranged structural features of the electrolyte contribute to these correlations. This lack of clear connection between structural properties and transport presents substantial challenges in understanding the molecular origins of conductivity and ultimately designing electrolytes with optimal transport properties \cite{balos2020macroscopic}.

A common way to intuitively relate local structure and global dynamics in electrolytes is by introducing the concept of ion pairing and clustering \cite{doi:10.1021/cr040087x, PhysRevLett.122.136001, https://doi.org/10.1002/cphc.202000153, MAIER201421}. This builds upon Arrhenius's work, conducted over a century ago, which used conductivity to provide evidence for the existence of ion association in dilute electrolytes \cite{arrhenius1887dissociation}. The idea is that ions that are paired, or clustered, will contribute differently to the overall conductivity of the system than if they were free, and a reduction in conductivity can be explained by an increase in the number of neutral ion pairs. The ion pair picture can be extended by the ``cluster Nernst-Einstein" approach \cite{PhysRevLett.122.136001} (Figure \ref{fig:schematic}B), which considers each ionic aggregate as a distinct charge carrier whose net charge and self-diffusion coefficient are used in the Nernst-Einstein equation to predict the conductivity. However, recent work demonstrated that the static picture of ion pairing (defined using a distance criterion alone) may be insufficient to explain observed transport properties in many non-dilute electrolytes \cite{doi:10.1021/acs.macromol.0c02545}. Studies on polymer electrolytes \cite{shen_2020} and polyelectrolyte solutions \cite{fong2019ion,fong2020onsager}, for example, have found that the fraction of ions in pairs or aggregates can be anti-correlated with the extent of correlation in ions' motion, as quantified by the Onsager transport coefficients. These findings raise two key questions: First, can the dynamic behaviour of concentrated electrolytes be predicted from static, short-ranged ionic structure alone? If so, what are the ionic environments that contribute to high or low conductivity?

In this work, we will first show that machine learning on molecular dynamics simulation data successfully decomposes global conductivity into a spatio-temporal average of local, instantaneous ionic contributions, and relate this ``local molar conductivity" field to the local short-ranged ionic environment (Figure \ref{fig:schematic}C). This machine learning model is generalizable to unseen electrolyte systems, suggesting that dynamic behaviour of concentrated electrolytes can be predicted from static, short-ranged ionic environment alone. We will then show that mechanistic insights about ionic conductivity can be gleaned by analysing the spatial and temporal distribution of this machine-learnt local instantaneous conductivity field. Our analysis suggests that the suppression in molar conductivity in concentrated electrolytes is not due to distinct ion pairs, but a long heavy tail of less conductive ions, themselves spatially correlated, which emerges due to like-charge interactions. 

\section*{Results}

\subsection*{Machine-learnt local instantaneous conductivity}
We employ machine learning to decompose the system conductivity into local ionic contributions. We posit there exists some mapping from the local environment $\mathbf{x}_i(t)$ of ion $i$ at time $t$ (accessible from molecular dynamics simulations), to an instantaneous local molar conductivity $k_i(t) = f_{\theta}(\mathbf{x}_i(t))$, such that the molar conductivity $\Lambda$ is the time and system average of the contributions from all ions in the system:
\begin{equation}
    \Lambda = \frac{ 1}{NT}\sum\limits_{t=1}^T\sum\limits_{i=1}^N k_i(t),
\end{equation}
where $N$ is the number of ions and $T$ is the number of temporal snapshots. Note that the existence of a function $k_i(t)$ is our key hypothesis, which remains to be tested. 

To test whether $k_i(t)$ exists, we use a machine learning approach. We parameterise $k_i(t) = f_{\theta}(\mathbf{x}_i(t))$, where $\theta$ are parameters that we infer from simulations by minimising the weighted loss function
\begin{equation}
    \mathcal{L}(\theta) = \sum\limits_{s=1}^S \frac{\left(\Lambda^{(s)} - \frac{ 1}{NT}\sum\limits_{t=1}^T\sum\limits_{i=1}^N f_{\theta}(\mathbf{x}^{(s)}_i(t)) \right)^2}{\Delta^{(s)}},
\end{equation}
where $\Lambda^{(s)}$ and $\Delta^{(s)}$ are the molar conductivity and variance in molar conductivity of system $s$, respectively. If $\mathcal{L}(\theta)$ can be minimised, and the resulting $k_i(t)$ can accurately predict the molar conductivity of unseen electrolyte systems, then we have some confidence in the validity of the concept of local conductivity. 

\subsection*{Local structure predicts ionic conductivity of electrolytes}
\begin{figure}
\centering
\includegraphics[width=.4\linewidth
, trim={0cm, 0cm, 0cm, 0}, clip]{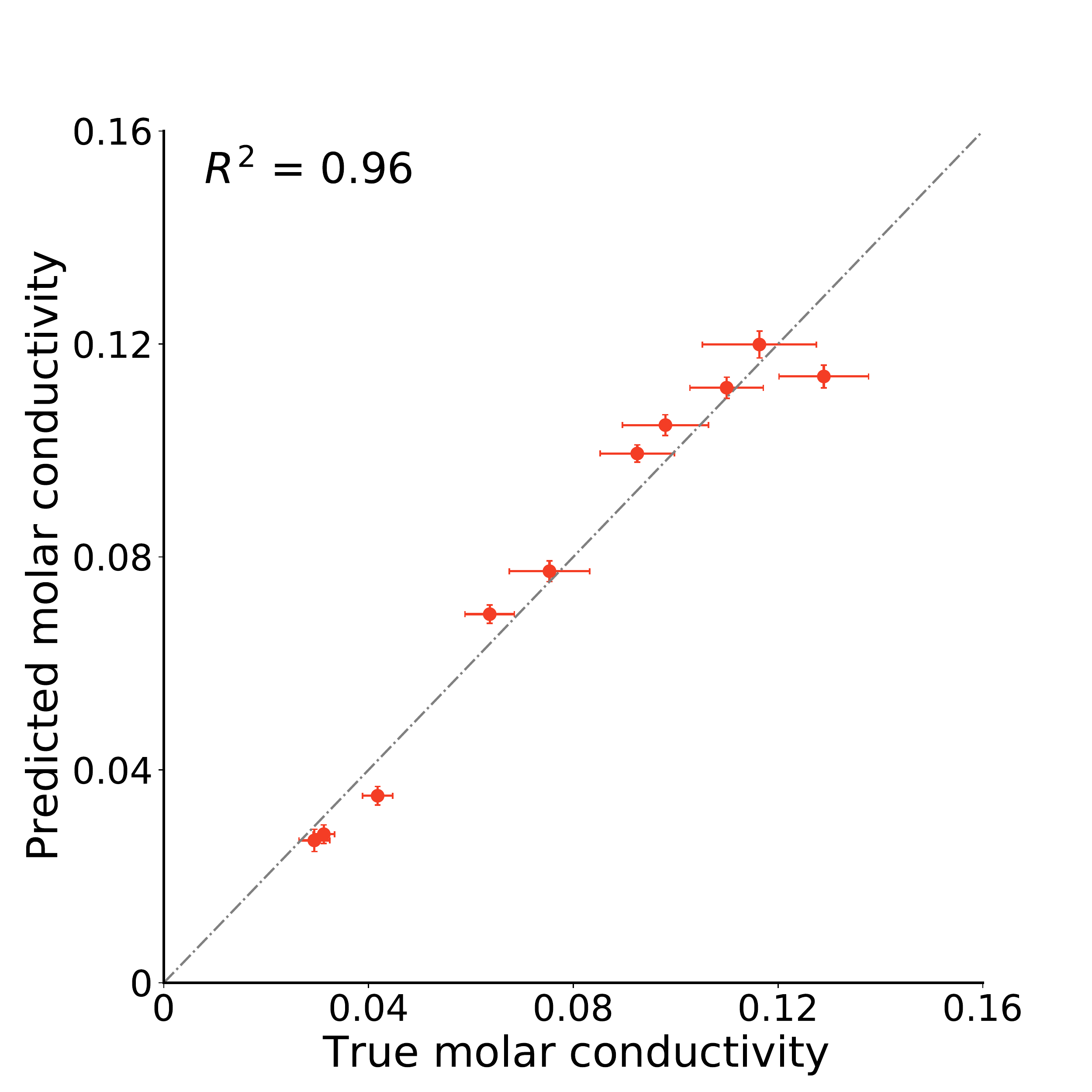}
\caption{Our model accurately predicts the molar conductivity of held out test systems, given knowledge only of the local environments inhabited by the ions in that system.}
\label{fig:parity}
\end{figure}

We begin by testing whether a model which is trained on simulation data from a very simple electrolyte can accurately predict the behaviour of held out test systems from the same class of electrolytes, but different concentrations and Bjerrum lengths. The model herein consists of charged and neutral Lennard-Jones particles to represent the ions and solvent, respectively (see Materials and Methods).

Figure \ref{fig:parity} shows that the model accurately predicts molar conductivity of each unseen electrolyte system, attaining an $R^2$ score of $0.96$. This is at the upper bound of attainable accuracy of $R^2 = 0.96 \pm 0.02$ which is constrained by the uncertainty in the molar conductivity values estimated using Green-Kubo relations. We note that the model only incorporates short ranged interactions between ions (up to 5.0$\sigma$). Therefore, these results appear to suggest that molar conductivity of electrolytes can be predicted given local static structure alone, although electrostatic forces themselves are long ranged. However the relevant local structure is not necessarily ion pairs. 

We next seek to determine the extent to which this model is physically meaningful, by testing whether it can generalise to a different class of electrolytes without further retraining. We achieve this by simulating an additional sixteen electrolytic systems, but now some fraction of ions are permanently bonded to an unlike charge via a harmonic bond potential, to form permanent ion pairs. These permanent pairs contribute to dielectric screening and provide polar solvation to neighbouring ions. Because neutral ion pairs do not contribute to overall system conductivity, we do not compute a local conductivity for ions in these pairs. Crucially however, the local environments of each of the non-harmonically bonded ions accounts for the presence of the ion pairs in its own local environment representation. We look at systems with low (0.0005$\sigma^{-3}$ / 2.5$\sigma$), medium (0.01$\sigma^{-3}$ / 4.0$\sigma$) / (0.025$\sigma^{-3}$, 6.0$\sigma$) and high (0.05$\sigma^{-3}$ / 10.0$\sigma$) concentrations and Bjerrum lengths, and for each system we consider the case where 10\%, 20\%, 30\% and 40\% of the ions are permanently bonded to an unlike charge. 

\begin{figure}[ht]
\centering
\includegraphics[width=.4\linewidth, trim={0cm, 0cm, 0cm, 0}, clip]{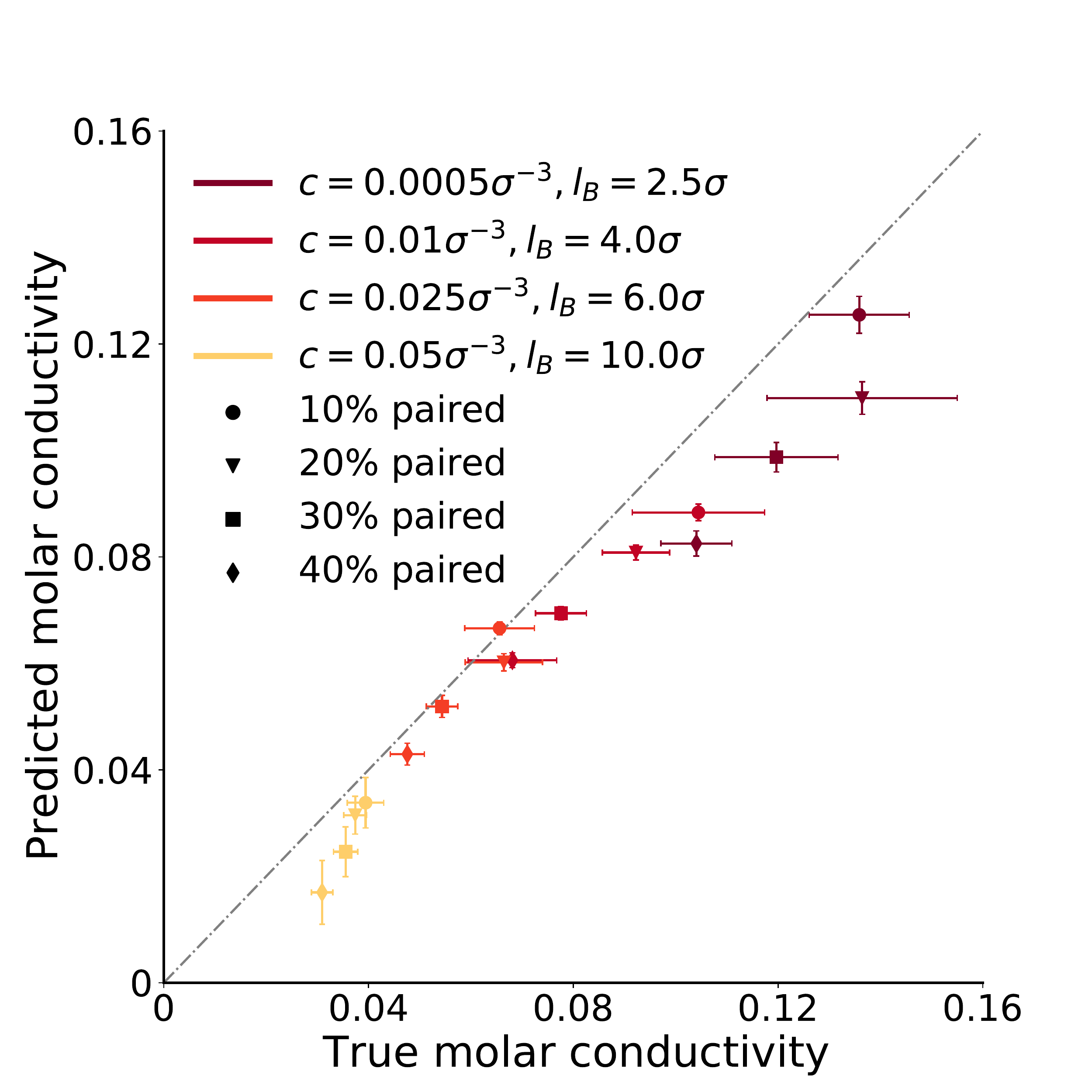}
\caption{\textbf{Testing for external validity.} The model can predict the molar conductivity of systems which now contain a proportion of permanently bonded ion pairs. The model was not trained using data from any such system and so is `extrapolating' outside of the training domain. }
\label{fig:parity-ion-pairs}
\end{figure}

Figure \ref{fig:parity-ion-pairs} shows that the model, trained only on non-polar solvent, retains a high accuracy when extrapolating into this new class of systems in which new physics are present (i.e. the existence of harmonic potentials causing permanent ion pairs/dipoles in the system). This suggests that the model is generalisable to qualitatively different electrolyte systems. A strength of the model is that we effectively transform an extrapolation problem into an interpolation problem since local environments that are detected in the systems with no permanent ion pairs also exist (to differing degrees) in the systems with permanent ion pairs.

\subsection*{Heterogeneity in local conductivity correlates with system molar conductivity}

\begin{figure}[ht]
    \centering
\includegraphics[width=0.9\linewidth, trim={0cm, 41cm, 0cm, 30cm}, clip]{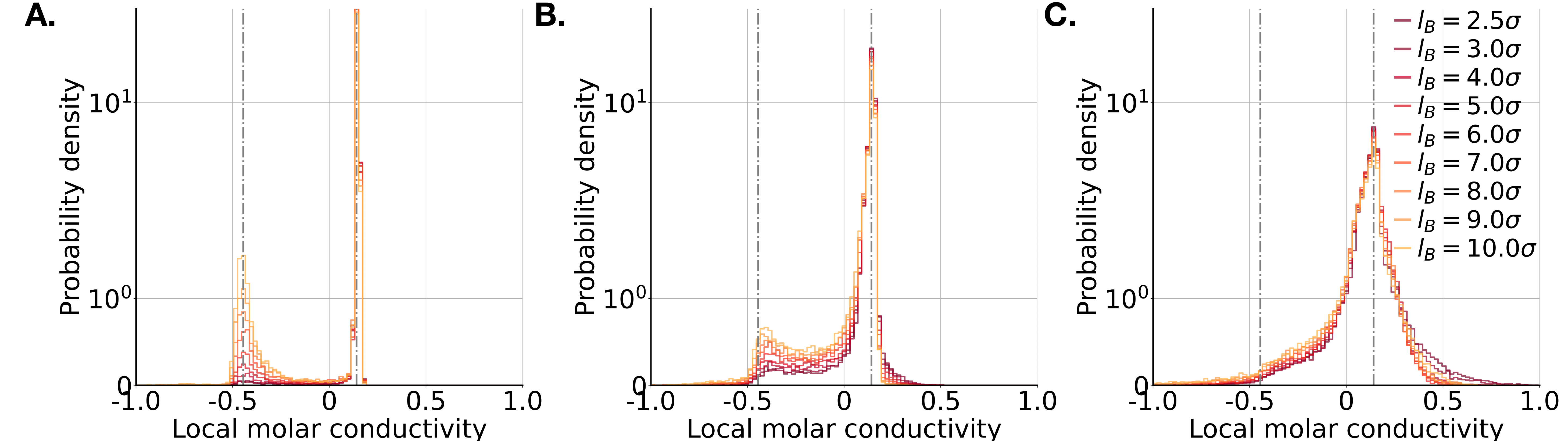}
    \caption{\textbf{Local molar conductivity distribution.}\textbf{ A.} At low concentrations (0.0005$\sigma^{-3}$) the local conductivity distribution is bi-modal. \textbf{B.} As the concentration increases (to 0.01$\sigma^{-3}$), these modes gradually become less distinct. \textbf{C.} At high concentrations (0.05$\sigma^{-3}$), the modes can no longer be distinguished.}
    \label{fig:histograms}
\end{figure}

Having shown that local structure can predict dynamics, we now seek to distil a mechanistic understanding of ion transport using this machine-learnt concept of localised molar conductivity. 

We begin by analysing the distribution of local molar conductivities and how this changes as the concentration and Bjerrum length increase. Figure \ref{fig:histograms}A shows that at low concentrations the distribution of local conductivities is bi-modal, with the majority of particles contributing positively to overall ion conduction whilst a minority of environments inhibit ion conduction and have a negative local conductivity. For a fixed concentration, the proportion of ions negatively contributing to conductivity increases as the Bjerrum length increases. Ions negatively contributing to molar conductivity tend to be closer to an unlike charge -- at the lowest concentration of 0.0005$\sigma^{-3}$ and Bjerrum length 2.5, 22.6\% of ions with unlike charge located within the first solvation shell (at a distance less than 1.6$\sigma$) are predicted to have a negative conductivity, whilst
just 0.07\% of ions farther than 1.6$\sigma$ from any unlike charge are predicted to contribute negatively to molar conductivity. 

On the surface, this might suggest the ion pair picture, defined using inter-ionic distance, explains ionic conductivity. However, as the concentration increases (Figures \ref{fig:histograms}B and \ref{fig:histograms}C), the distribution becomes uni-modal, albeit with a heavy tail skewing towards ionic environments that inhibit conductivity. The lack of a bi-modal distribution suggests that looking at discrete ion pairs does not provide the full picture, and is in agreement with prior findings that at high concentrations there are not multiple statistically distinct types of ionic environment \cite{doi:10.1063/5.0039617}. This is further supported by the much more comparable fractions of `clustered' and `unpaired' ions that contribute negatively to molar conductivity -- at a concentration of 0.05$\sigma^{-3}$ and Bjerrum length of 10.0$\sigma$, 28.0\% of ions with unlike charge located within the first solvation shell (at a distance less than 1.6$\sigma$) are predicted to have a negative conductivity, whilst
21.6\% of ions farther than 1.6$\sigma$ from any unlike charge are predicted to contribute negatively to molar conductivity. 

\begin{figure}[ht]
    \centering
\includegraphics[width=.4\linewidth]{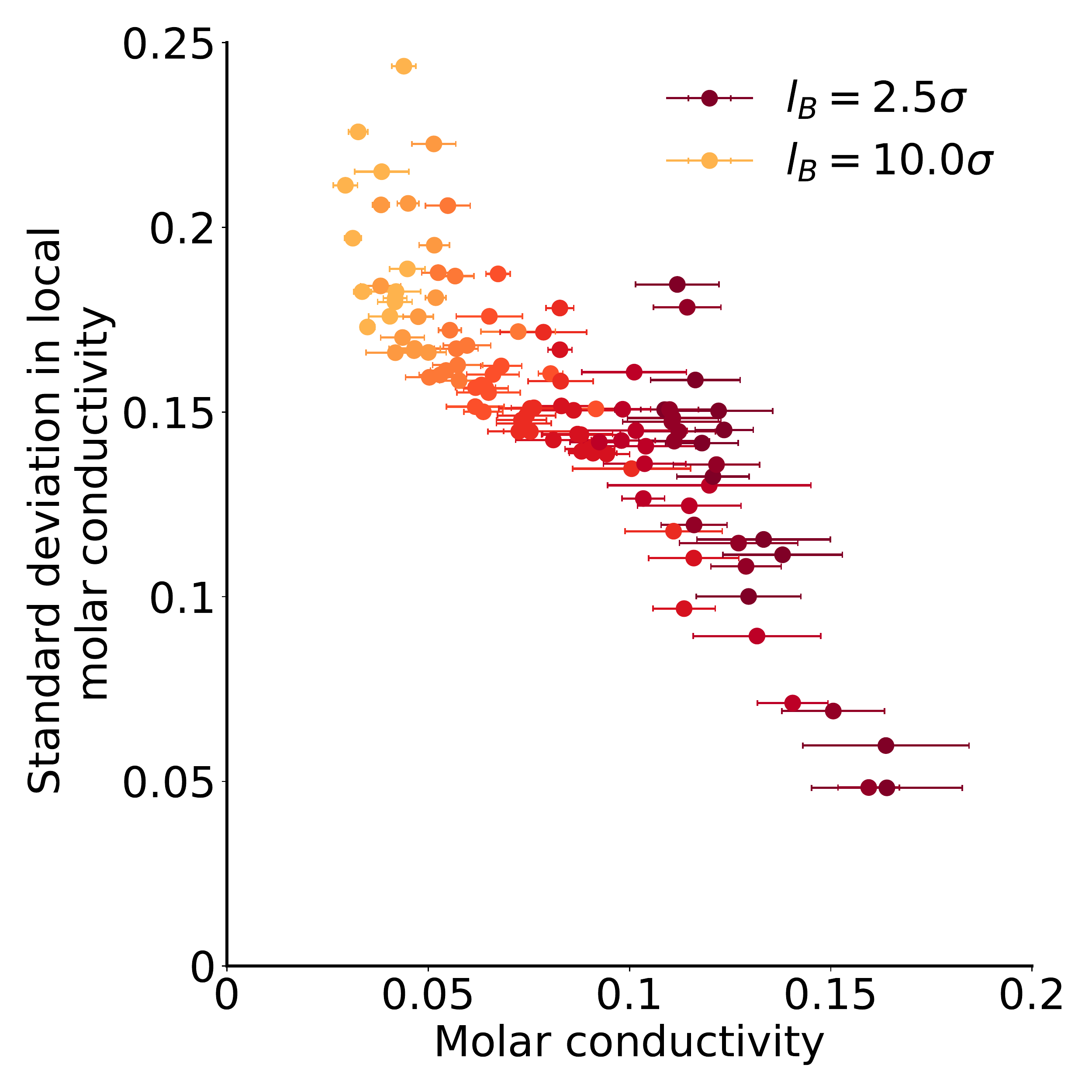}
    \caption{\textbf{Heterogeneity in local molar conductivity.} The standard deviation of the localised molar conductivity distribution is negatively correlated with the collective molar conductivity.}
    \label{fig:fluctuations}
\end{figure}

The existence of a heavy tail in the conductivity distribution motivates us to explore the variance of the local conductivity as an organising principle. Figure \ref{fig:fluctuations} shows that low molar conductivity is correlated with high variance in local conductivity. In other words, slow ion transport appears to be associated with heterogeneity in local conductivity.  Surprisingly, despite there being a bi-modal distribution at low concentrations and a uni-modal distribution at high concentrations, the variance of the local conductivity distribution is greater at higher concentrations. 

\subsection*{Local conductivity is spatially correlated}

\begin{figure}[ht]
    \centering
\includegraphics[width=0.9\linewidth, trim={0cm, 26cm, 0cm, 8cm}, clip]{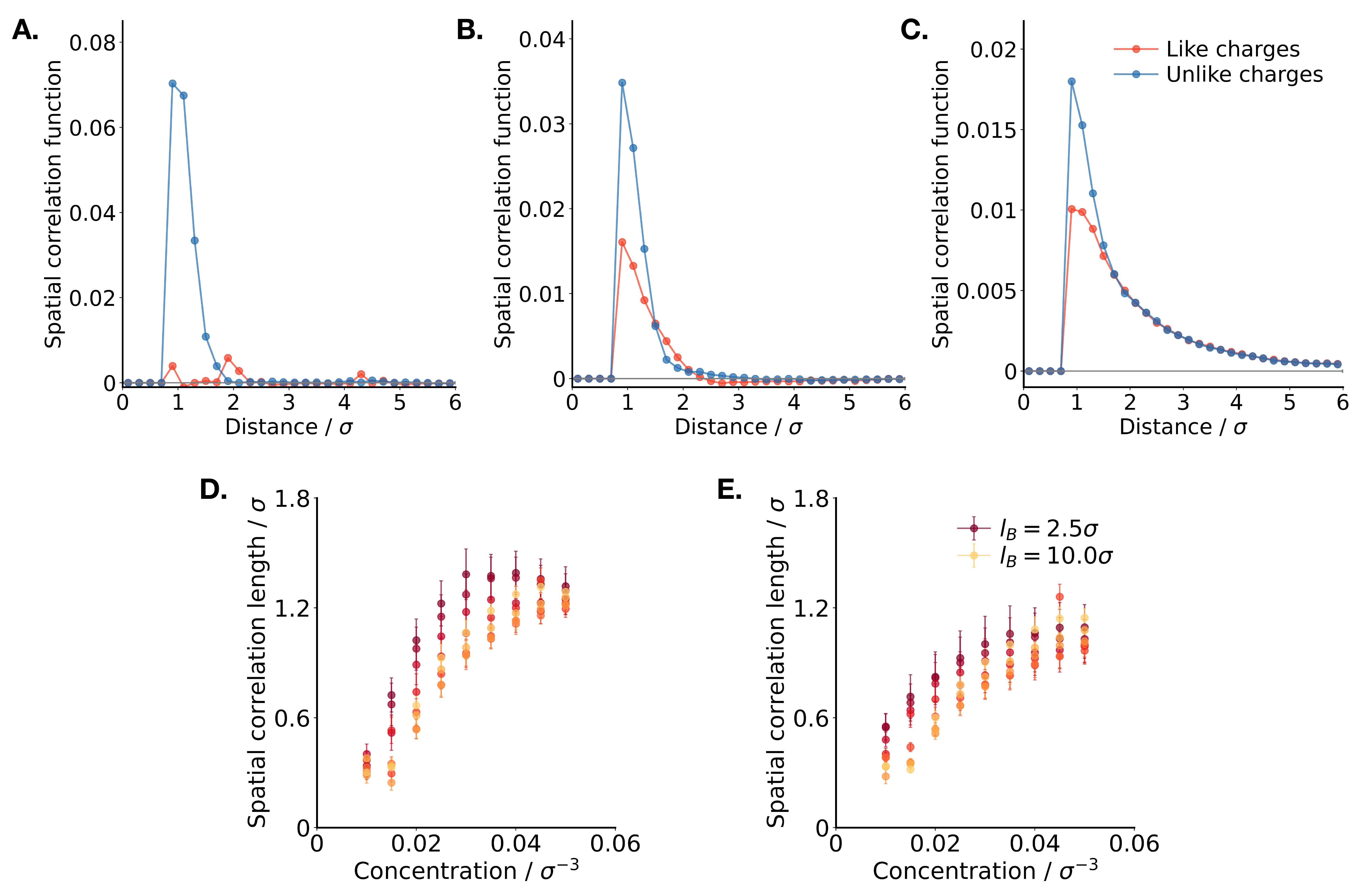}
    \caption{\textbf{The local conductivities of neighbouring like and unlike charges are positively spatially correlated.} Figures \textbf{A}, \textbf{B} and \textbf{C} show the like and unlike charge spatial correlation functions for concentrations 0.0005$\sigma^{-3}$, 0.01$\sigma^{-3}$ and 0.05$\sigma^{-3}$, and Bjerrum lengths 2.5$\sigma$, 6.0$\sigma$ and 10.0$\sigma$, respectively. The magnitude of correlations decreases with inter-particle distance, and can be fitted to an exponential decay function to extract two correlation lengthscales. Figures \textbf{D} and \textbf{E} show how the like and unlike charge correlation lengthscale changes with concentration, respectively.}
    \label{fig:spatial-correlations}
\end{figure}

We next analyse the spatial correlations in local conductivity, and decompose this correlation into like-charge and unlike-charge contributions. We define the spatial correlation function $C(r)$ between species $A$ and $B$ as
\begin{equation}
C(r) = \frac{\sum\limits_{t=1}^T\sum\limits_{i=1}^{N_A}\sum\limits_{j \in |r_{ij}(t) - r|<\delta/2}\Delta k^{(A)}_i(t)\Delta k^{(B)}_j(t)}{\sum\limits_{t=1}^T \sum\limits_{i=1}^{N_A} \sum\limits_{j \in |r_{ij}(t) - r|<\delta/2} 1},
\end{equation}

where $T$ is the number of snapshots, $N_A$ is the number of ions of type $A$, $r_{ij}(t)$ is the distance between particles with conductivities $k^{(A)}_i(t)$ and $k^{(B)}_i(t)$ at snapshot $t$, and $\Delta k^{(A)}_i(t)=k^{(A)}_i(t) - \bar{k}^{(A)}(t)$ where $\bar{k}^{(A)}(t)$ is the average conductivity of ions of type $A$ at snapshot $t$. $\delta$ is the size of the histogram bin and set to 0.2$\sigma$. The denominator measures the number of occurrences when type $A$ ions and type $B$ ions are separated by a distance in the range $\{r-\delta/2, r+\delta/2\}$. This enables us to ask: if two particles of types $A$ and $B$ are observed to be at a distance $r$ from each other, how correlated are their local conductivities on average? 

Figure \ref{fig:spatial-correlations} shows that the local conductivities of neighbouring like and unlike charges are positively spatially correlated, with the correlations decaying as a function of inter-particle distance. This suggests that conductivity can be spatially localised to some extent. At low concentrations (Figure \ref{fig:spatial-correlations}A) the dominant spatial correlations are at short distances between unlike charges, which can be attributed to the formation of neutral ion pairs. However as the concentration and Bjerrum length increase (Figure \ref{fig:spatial-correlations}B), the magnitude of like charge correlations increases relative to unlike charge correlations, such that at the highest concentrations (Figure \ref{fig:spatial-correlations}C), the like charge local conductivity correlations are almost as large as the unlike charge correlations.

The lengthscale of the correlation can be quantified by fitting to an exponential decay function. We observe that the spatial correlation lengthscale increases as concentration increases, for both neighbouring like charges (Figure \ref{fig:spatial-correlations}D) and unlike charges (Figure \ref{fig:spatial-correlations}E). As such, there appears to be a change in transport mechanism from short range ordering at very low concentrations to longer range collective motion at higher concentrations. 

\subsection*{Relating local conductivity to global dynamic properties}
Our analysis thus far provides spatial resolution to conductivity in electrolytes. However, a question is whether our machine-learnt notion of local instantaneous conductivity can be related to dynamic quantities in non-equilibrium statistical physics. Such a relation will provide further evidence that the local conductivity represents a physically meaningful way to interpret electrolyte transport. As such, we analyse how the local conductivity field is related to the Onsager coefficients for each system. The Onsager formalism \cite{PhysRev.37.405, PhysRev.38.2265} can help to bridge the gap between molecular level and macroscopic scale analysis of ion transport \cite{doi:10.1021/acs.macromol.0c02545, https://doi.org/10.1002/aic.17091}, with each Onsager coefficient characterising the extent to which motion of two species is correlated on average. We focus here on two Onsager coefficients, firstly, $L^{++}_\mathrm{self}$, which captures ideal uncorrelated motion of ions ($L^{++}_\mathrm{self}$ is related to the cation self-diffusion coefficient $D_+$ by $L^{++}_\mathrm{self} = D_+ c / RT$, where $c$ is the salt concentration and $RT$ is the thermal energy), and secondly $L^{+-}$, which captures the correlation between all anions and cations. 

\begin{figure}[ht]
    \centering
\includegraphics[width=.4\linewidth]{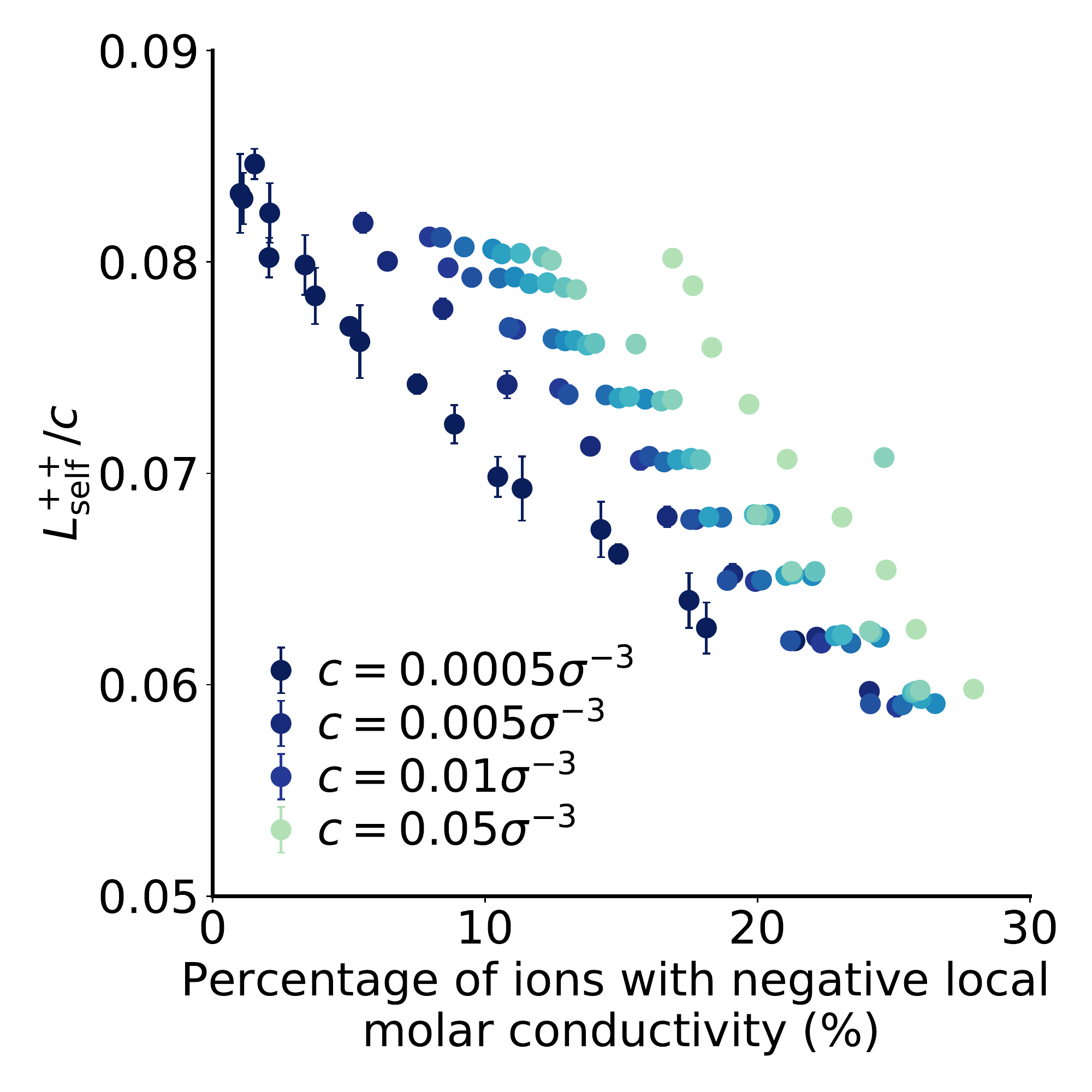}
    \caption{\textbf{Relation between the local molar conductivity field and ideal, uncorrelated motion.} The fraction of ions whose local molar conductivity is negative is related to the amount of ideal, uncorrelated motion (as quantified by $L^{++}_\mathrm{self} / c$).}
    \label{fig:onsager-self}
\end{figure}

\begin{figure}[ht]
    \centering
\includegraphics[width=.4\linewidth]{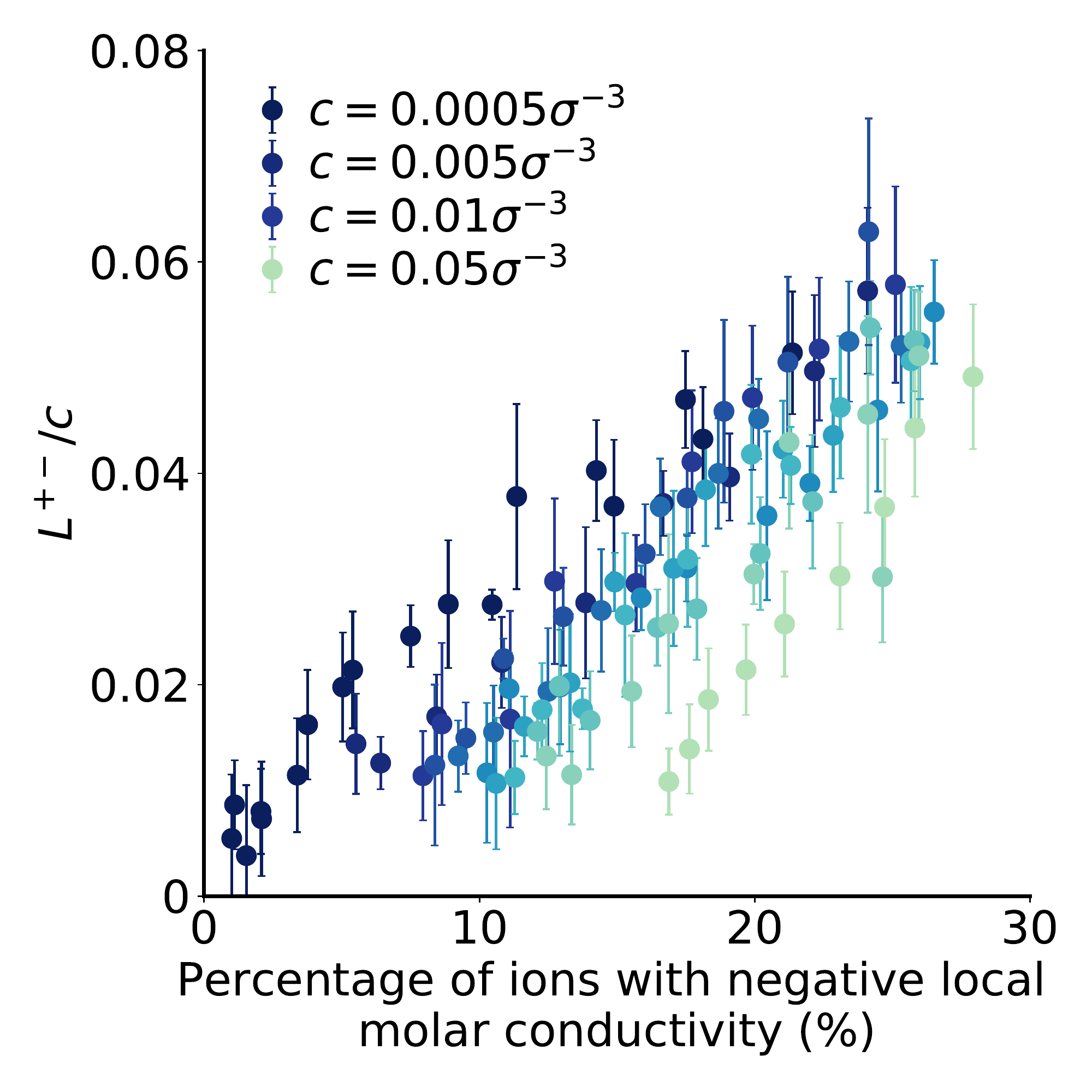}
    \caption{\textbf{Relation between the local molar conductivity field and correlated cation-anion motion.} The fraction of ions whose local molar conductivity is negative is positively correlated with the magnitude of cation-anion correlations (as quantified by $L^{+-}/ c$). This is in line with the fact that strong cation-anion correlations negatively impact a system's conductivity \cite{doi:10.1021/acs.macromol.0c02545}.}
    \label{fig:onsager-cross}
\end{figure}

Figure \ref{fig:onsager-self} shows the correlation between $L^{++}_\mathrm{self}/c$ and the percentage of ions with negative local conductivity. We observe that for a given concentration, as the proportion of ions with negative local conductivity decreases, $L^{++}_\mathrm{self}/c$ (and hence $D_+$) increases. This makes intuitive sense as we would generally expect ions with inhibited motion to contribute less positively to molar conductivity. Additionally, in Figure \ref{fig:onsager-cross} we observe a strong positive correlation between the proportion of ions with negative local conductivity and $L^{+-}$. This positive correlation is in line with the fact that positive $L^{+-}$ values decrease overall ionic conductivity. The intuitive connection between $L^{+-}$ and the local molar conductivity stands in contrast with the relationship between $L^{+-}$ and the static picture of ion pairing: in previous works, it has been observed that at high concentrations ion motion is less correlated (corresponding to a reduction in $L^{+-}$), despite an increase in the proportion of ions that are paired \cite{shen_2020}. This observation that the fraction of ion pairs is not correlated with the ionicity suggested that ion pairing cannot explain the extent of cation-anion correlations that determine transport properties at non-dilute concentrations, and that ion pair dynamics are much more correlated with cation-anion correlations than statics \cite{doi:10.1021/acs.macromol.0c02545}. However, the local conductivity, computed using local statics alone, appears to capture just as much physics as using a dynamic picture of ion pairing. 

\section*{Discussion}
In this paper we have demonstrated that it is possible to predict the molar conductivity of a simple electrolyte model, from short-ranged static structure alone, using machine learning. Our model generalises without retraining to a different class of electrolytes with dipolar screening, suggests that it is at least to some extent physically meaningful. 

A by-product of our model is a framework that delivers insights into the individual atomistic contribution that each ion, instantaneously experiencing a particular local environment, makes to macroscopic conductivity. This allows us to explore the microscopic determinants of conductivity. We find that at low concentrations, there are two statistically distinct modes of conduction, with the majority of ions contributing positively to ion conduction whilst a small minority inhibit conduction. These modes are found to be related to conventionally defined `free' and `paired' states. In contrast, at high concentrations, the distribution of local molar conductivities broadens and becomes uni-modal, suggesting that it is not statistically sound to model concentrated electrolytes as comprising well defined clusters of ions. In addition, we observe that systems with low variance in local molar conductivity exhibit higher total molar conductivity in general, and local molar conductivity is spatially correlated through both like and unlike-charge correlations.

More broadly, our approach demonstrates that machine learning can be a fruitful tool for unravelling new insights about soft matter systems, by attributing collective properties to individual atomistic contributions.

\section*{Materials and Methods}
\subsection*{Molecular dynamics simulations}

A coarse-grained molecular dynamics (MD) simulation model was used to simulate bulk electrolyte systems with a range of concentrations and Bjerrum lengths ($l_\mathrm{B}$). All simulations were performed using the HOOMD-blue package \cite{anderson2020hoomd}. In total, 108 systems were simulated with twelve concentrations ranging from 0.0005 $\sigma^{-3}$ to 0.05 $\sigma^{-3}$ and nine Bjerrum lengths in the range 2.5 $\sigma$ to 10.0$ \sigma$. Each simulation consisted of cations, anions, and solvent, all of which are modelled as beads of diameter $\sigma$ (the Lennard-Jones unit of distance) and unit mass. If the size of each bead is mapped to the size of a water molecule (2.75 \AA), the concentration range of 0.0005 $\sigma^{-3}$ to 0.05 $\sigma^{-3}$ approximately corresponds to 0.04 M to 4 M in real units. All particles interact via a truncated, shifted Lennard-Jones (LJ) potential:
\begin{equation}
    U_{\mathrm{LJ}} (r) = 
    \begin{cases} 
      4\varepsilon\bigg[\big(\frac{\sigma}{r}\big)^{12}-\big(\frac{\sigma}{r}\big)^{6}-\big(\frac{\sigma}{r_{\mathrm{cut}} }\big)^{12}+\big(\frac{\sigma}{r_{\mathrm{cut}} }\big)^{6}\bigg] & r\leq r_{\mathrm{cut}}  \\
      0 & r> r_{\mathrm{cut}} 
   \end{cases}~,
\end{equation}
where $\varepsilon$ is the LJ unit of energy and the cutoff distance $r_{\mathrm{cut}}$ is chosen to be $2^{1/6}\sigma$. Additionally, cations and anions interact via a Coulomb potential,
\begin{equation}
    U_{\mathrm{Coulomb}}(r) = k_{\mathrm{B}}T \frac{l_\mathrm{B}z_i z_j}{r}~,
\end{equation}
where $k_\mathrm{B} T$ is the thermal energy and $z_i$ is the charge of species $i$ (set to $+1$ for cations and $-1$ for anions). Long-ranged Coulombic interactions were computed using the PPPM method \cite{lebard2012self}.

Simulation cells were prepared by randomly packing all particles into a cubic simulation box using PACKMOL \cite{Martinez2009} at a density of 0.8 $\sigma^{-3}$. A standard simulation consisted of 40,000 total particles, with the relative quantities of ions and solvent varied to adjust the salt concentration. Six independent replicates were prepared for each concentration/Bjerrum length combination, one of which included 80,000 total particles to verify that the simulation results were not influenced by finite size effects. The as-prepared systems were initially equilibrated using the Fast Inertial Relaxation Engine (FIRE) minimisation algorithm \cite{bitzek2006structural}. Simulations were run with a time step of 0.005 $\tau$ (the LJ unit of time) for a total of $10^7$ steps. The temperature of each simulation was maintained at $k_B T / \varepsilon = 1$ using a Nos\'e-Hoover thermostat. 

The Onsager transport coefficients ($L^{ij}$) of each system were computed using the following Green-Kubo relations \cite{fong2020onsager,https://doi.org/10.1002/aic.17091,doi:10.1021/acs.macromol.0c02545}:
\begin{equation}\label{lij}
    L^{ij} = \frac{1}{6k_{\mathrm{B}}TV}\lim_{t\to\infty}\frac{d}{dt} \big< \sum\limits_{\alpha}[\boldsymbol{r}_i^{\alpha}(t)-\boldsymbol{r}_i^{\alpha}(0)]\cdot \sum\limits_{\beta}[\boldsymbol{r}_j^{\beta}(t)-\boldsymbol{r}_j^{\beta}(0)]\big>~,
\end{equation}
where the indices $i$ and $j$ refer to each type of species (cation, anion, and solvent) and the indices $\alpha$ and $\beta$ refer to individual particles. $V$ is the system volume and $\boldsymbol{r}_i^{\alpha}$ is the position of particle $\alpha$ (of type $i$) relative to the center-of-mass position of the entire system. The self transport coefficients, $L^{ii}_{\mathrm{self}}$, which are related to the self-diffusion coefficients of a given species via $ L^{ii}_{\mathrm{self}} = \frac{D_i c_i}{k_{\mathrm{B}}T}$, may be computed similarly using
\begin{equation}\label{l_ii_self}
    L^{ii}_{\mathrm{self}} = \frac{1}{6k_{\mathrm{B}}TV}\lim_{t\to\infty}\frac{d}{dt} \sum\limits_{\alpha}\big< [\boldsymbol{r}_i^{\alpha}(t)-\boldsymbol{r}_i^{\alpha}(0)]^2\big>~.
\end{equation}
In implementing \eqref{lij} and \eqref{l_ii_self}, it is ensured that the terms in angular brackets exhibit at least a decade of linearity, i.e., the systems are safely in the diffusive regime. From $L^{ij}$, the total ionic conductivity $\kappa$ may be computed according to
\begin{equation}\label{conductivity}
    \kappa = F^2\sum\limits_i \sum\limits_j L^{ij}z_iz_j ~,
\end{equation}
where $F$ is Faraday's constant. Error bars for $\kappa$ and $L^{ij}$ are given as the standard deviation of six independent simulations.

\subsection*{Machine learning model}
We assume there exists some mapping from the local environment $\mathbf{x}_i(t)$ of ion $i$ at time $t$, to instantaneous local molar conductivity $k_i(t)$, such that $k_i(t) = f_{\theta}(\mathbf{x}_i(t))$. The instantaneous molar conductivity of each species is the average of the local molar conductivities, i.e. 
\begin{equation}
    \Lambda_{\pm}(t) = \frac{1}{N_{\pm}} \sum\limits_{i = 1}^{N_{\pm}} k_{i\pm}(t).
\end{equation}
The global molar conductivity, $\Lambda$, is the average over time, and across ion species, such that 
\begin{equation}
    \Lambda = \frac{1}{2}(\Lambda_{+} + \Lambda_{-}) = \frac{1}{2T} \sum\limits_{t=1}^T (\Lambda_{+}(t) + \Lambda_{-}(t)), 
\end{equation}
where $T$ is the number of snapshots. 

In order to ensure that the structural environments present in each snapshot are independent from the others, the sampling interval of the molecular dynamics trajectories was chosen to be longer than the average ion pair lifetime in the system. The molar conductivity $\Lambda$ is then
\begin{equation}
    \Lambda = \frac{1}{NT}\sum\limits_{t=1}^T\sum\limits_{i=1}^{N} k_i(t),
\end{equation}
where $N = N_+ + N_-$ is the total number of ions in the system.

We learn the mapping $k_i(t) = f_{\theta}(\mathbf{x}_i(t))$ by minimising the weighted loss function
\begin{equation}
    \mathcal{L}(\theta) = \sum\limits_{s=1}^S \frac{\left(\Lambda^{(s)} - \frac{ 1}{NT}\sum\limits_{t=1}^T\sum\limits_{i=1}^N f_{\theta}(\mathbf{x}^{(s)}_i(t)) \right)^2}{\Delta^{(s)}},
\end{equation}
where $\Lambda^{(s)}$ and $\Delta^{(s)}$ are the molar conductivity and variance in molar conductivity of system $s$, respectively. For each system, at each time $t$ there are $N$ measured local ionic environments $\{\mathbf{x}_i(t)\}_{i=1}^N$ such that the total number of measured local environments (and corresponding local molar conductivities) per system is $NT$.

In practice, we approximate the function $f_{\theta}(\mathbf{x}_i(t))$ by training an ensemble of 25 neural networks each with two hidden layers of dimensionality 100. We split the data using a random 90\%/10\% train-test split and then split the training set into five parts. We use five-fold cross validation, and train five neural networks per train/validation random split, each with a different initialisation seed. The network parameters $\theta$ are optimised using the Adam optimiser \cite{https://doi.org/10.48550/arxiv.1412.6980} for up to 1000 epochs at a learning rate of 0.0005. We adopt the early stopping technique to limit the risk of overfitting, by stopping training when the RMSE on the validation dataset is minimised. 

\subsection*{Representation of local environment}
In principle, $\mathbf{x}_i(t)$ can be any representation of the ionic environments. To test specifically whether short-ranged local structure predicts conductivity, we implement the Smooth Overlap of Atomic Positions (SOAP) framework \cite{PhysRevB.87.184115}, using the DScribe package \cite{HIMANEN2020106949}, to form a rotationally and translationally invariant vectorised representation of each ion's local environment. Here, to determine the local environment of some central particle, we consider all neighbouring particles and the central particle to be represented by a Gaussian distribution. Then we can independently consider the contribution of each particle species, with the contribution from type $Z$ particles to the local environment of central particle $a$ being
\begin{equation}
    \rho^{(Z)}_a(\mathbf{r}) = \sum_{\substack{z \in Z \\|\mathbf{r}_{az}| < r_C}}\exp\left(-\frac{|\mathbf{r}_{az} - \mathbf{r}|^2}{2\sigma^2}\right),
\end{equation}
where $\mathbf{r}_{az}$ is the displacement from the central particle to neighbouring particle $z$, $\sigma$ is the particle diameter, and we only consider the contribution from neighbouring particles closer than a cut-off distance $r_C = 5.0 \sigma$. This function is then rewritten as orthonormal spherical Gaussian type radial basis functions 
\begin{equation}
    g_{nl}(r) = \sum_{n'=1}^{n_{max}} \beta_{nn'l}r^l\exp({-\alpha_{n'l}r^2}),
\end{equation}
as well as a set of orthonormal spherical harmonic functions
\begin{equation}
    Y_{lm}(\theta, \phi) = \sqrt{\frac{(2l+1)(l-m)!}{4\pi (l+m)!}}P_l^m(\cos{\theta})\exp({im\phi}),
\end{equation}
such that 
\begin{equation}
    \rho^{(Z)}(\mathbf{r}) = \sum_{nlm}c_{nlm}^{Z}g_{nl}(r)Y_{lm}(\theta, \phi).
\end{equation}
Here, $\beta_{nn'l}$ and $\alpha_{n'l}$ are pre-determined weights, $P_l^m(\cos{\theta})$ are the associated Legendre functions and $c_{nlm}^Z$ are computed coefficients determined by the local environment. The final vector representation is formed from the concatenation of elements of the partial power spectrum, 
\begin{equation}
    p^{ZZ'}_{nn'l} = \pi\sqrt{\frac{8}{2l+1}}\sum_{m}c^{Z}_{nlm}(c^{Z'}_{n'lm})^*,
\end{equation}
where we set $n_{max}=6$ and $l_{max}=5$. For this work, we do not explicitly include solvent interactions in the local environment representation.
\subsection*{Data and code availability}
The data used to generate these results is freely available at the public repository https://doi.org/10.5281/zenodo.6957888. The code written to obtain these results is freely available at \url{https://github.com/PenelopeJones/conductivity}.

\begin{acknowledgments}
PKJ and AAL acknowledge the support of the Winton Programme for the Physics of Sustainability. PKJ acknowledges support from the Ernest Oppenheimer Fund. AAL acknowledges support from the Royal Society. KAP acknowledges the Joint Center for Energy Storage Research, an Energy Innovation Hub funded by the U.S. Department of Energy. KDF acknowledges support from the Berkeley Fellowship for Graduate Study. This research used the Lawrencium computational cluster resource provided by the IT Division at the Lawrence Berkeley National Laboratory (Supported by the Director, Office of Science, Office of Basic Energy Sciences, of the U.S. Department of Energy under Contract No. DE-AC02-05CH11231). This research also used resources provided by the Cambridge Service for Data Driven Discovery (CSD3) operated by the University of Cambridge Research Computing Service, provided by Dell EMC and Intel using Tier-2 funding from the Engineering and Physical Sciences Research Council (capital grant EP/P020259/1), and DiRAC funding from the Science and Technology Facilities Council (www.dirac.ac.uk).
\end{acknowledgments}

%

\end{document}